# Search for Supergalactic Anisotropies in the 3B Catalog

Dieter H. Hartmann[1], Michael S. Briggs[2], Karl Mannheim[3]

[1]*Dept. of Physics and Astronomy, Clemson University, Clemson, SC 29634*
[2]*Dept. of Physics, University of Alabama, Huntsville, AL 35899*
[3]*Universitäts-Sternwarte Göttingen, Geismarlandstrasse 11, D-37803 Göttingen*

The angular distribution of GRBs is isotropic, while the brightness distribution of bursts shows a reduced number of faint events. These observations favor a cosmological burst origin. If GRBs are indeed at cosmological distances and if they trace luminous matter, we must eventually find an anisotropic distribution of bright bursts. If a significant number of bursts originate at redshifts less than $z \sim 0.1$, the concentration of nearby galaxies towards the supergalactic plane is pronounced enough that we could discover the corresponding clustering of burst locations. We used the 3B catalog to search for a pattern visible in supergalactic coordinates. No compelling evidence for anisotropies was found. The absence of anisotropies in SG coordinates implies a minimum sampling distance of 200 $h^{-1}$ Mpc.

## INTRODUCTION

The isotropic angular distribution of $\gamma$-ray bursts (GRBs) argues in favor of their cosmological origin. However, this link only applies if bursts sample a significant fraction of the universe. On "small" scales the universe is known to be lumpy, and we expect local anisotropies in the angular pattern of GRB positions, but only if a significant number of bursts have occured close enough to trace these spatial inhomogeneities do we expect to find significant deviations from isotropy. The brightness distribution of bursts in the BATSE sample (8,12) suggests a maximum sampling redshift of $z_{max} \sim 1$, while the nearest bursts would originate at a redshift $z_{min} \sim 0.1$ (4,9,27). While the probability of observing bursts from distances less than a few 100 Mpc is thus apparently small, neither the extent of the local inhomogeneities nor the actual minimum sampling distance of BATSE are well know. It is thus useful to test the data for anisotropies of bright bursts.

If the burster brightness distribution indeed implies a sampling range $z \sim 0.1 - 1$, there would be too few bursts closer than $\sim 100$ Mpc, and we would not be able to detect local cosmic inhomogeneities. There are, however, a few indications that some bursts may have been much closer than the 100 Mpc limit suggested by their brightnesn distribution. Weak evidence for an association of ultra-high energy photons ($\sim 100$ TeV) with a GRB has been claimed for the bright burst GRB910511 (15). At distances in excess of D $\sim$ 300 Mpc such ultra-high energy photons or particles from bright bursts would not reach the Earth without excessive degradation through interaction with the cosmic microwave background or the intergalactic IR photons. The second indication is the possible association of GRBs with ultra-high energy cosmic rays (UHECRs) at E $\sim 10^{20}$ eV (13; but see 24-26). At 3 $10^{20}$ eV the particle horizon is less than 50 Mpc (7). While not conclusive, these observations suggest a closer origin of at least some GRBs, which motivates a search for nearby cosmic patterns.

On scales $\sim$ 100 Mpc the luminous universe is not uniform, but shows a well known concentration towards the supergalactic (SG) plane (5,6,18,19). Analysis of UHECR arrival directions suggests a concentration towards this plane (20), which may suggest that sources in the local universe are indeed responsible for some or all of the cosmic rays above $\sim 10^{19-20}$ eV (see also 1). We investigate the possibility that a significant number of GRBs may also originate in this inhomogeneous volume of the local universe. In any cosmological model where GRBs trace luminous matter, the anisotropies of the nearby universe must eventually be revealed.

## OBSERVATIONAL CONSTRAINTS

We use the 1122 improved BATSE positions of the 3B catalog (12) to test the angular distribution for a dipole or quadrupole moment in the supergalactic frame. We compile a subset of N = 867 bursts with measured peak fluxes. We calculate the dipole moment as the average

$$D_{SG} = \langle \cos \theta_{SG} \rangle \qquad (1)$$

where $\theta_{SG}$ is the angle between the direction to the GRB and the supergalactic center direction. Similarly, the quadrupole is

$$Q_{SG} = \langle \sin^2 SGB \rangle - 1/3 \qquad (2)$$

where SGB is the supergalactic latitude. If bursts are isotropic, the moments should average to zero with individual samples fluctuating according to a Gaussian distribution with $\sigma_D = 1/\sqrt{3N}$ and $\sigma_Q = 2/\sqrt{45N}$ (2). The non-uniform sky exposure induces artificial moments (3), which for $D_{SG}$ cause a bias of 0.022 and for $Q_{SG}$ a bias of 0.010. We normalize the difference between the observed values and the expected values (from isotropy plus uneven sampling) to the $1\sigma$ statistical uncertainties and quote these ratios as the intrinsic dipole or quadrupole deviation, $\Delta_D$ and $\Delta_Q$, in Table 1.

SG anisotropies are expected for nearby, bright sources. We assume a simple Friedmann universe with zero cosmological constant and standard candle GRBs. The fraction of bursts in the sample with redshifts less than z is f(z) = I(z,$\Omega$)/I($z_{max}$,$\Omega$), where

$$I(z, \Omega) = \int_0^z dz\ y(z)^2\ E(z)^{-1}\ (1+z)^{-1} \qquad (3)$$

with $E(z)$ and $y(z)$ defined in Peebles (14). To what redshift do we expect supergalactic anisotropies to be evident in the angular distribution of sources? The large scale structure of nearby radio sources (18,19) suggests strong SG concentrations to $z \sim 0.02$, corresponding to $D \sim 60\ h^{-1}$ Mpc, where h is the Hubble constant in units of 100 km s$^{-1}$ Mpc$^{-1}$. We shall assume $\Omega = 1$ hereafter. The maximum sampling distance is a crucial parameter. The standard interpretation of the log N − log P curve (9,27) yields $z_{max} \sim 1$. We thus expect local (within 100 Mpc) extragalactic structure to be detectable from a fraction $f \sim 1.4 \times 10^{-4}$ of the sample. It is thus very surprising that there may have already been coincidences of UHECRs with bright GRBs.

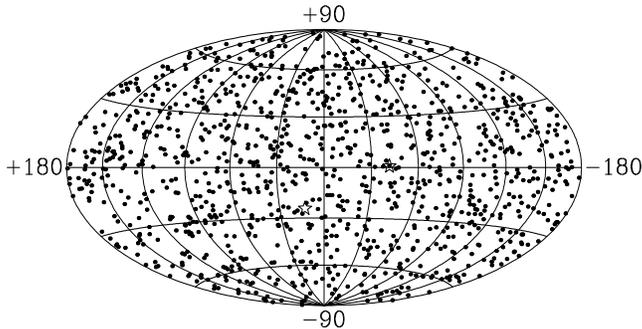

**FIG. 1.** Distribution of 3B bursts in supergalactic coordinates. Stars indicate two UHECRs (see text). The map is consistent with isotropy.

However, the sampling redshift of the BATSE detectors is not very well determined. Cohen & Piran (4) find that the maximum redshift could be as small as $z_{max} \sim 0.7$. Rutledge *et al.* (17) argue that that the maximum redshift could be as small as $z_{max} = 0.8$, while the most likely range is $1-2$. If we assume an extreme case of $z_{max} = 0.5$, the fraction of bursts originating within 100 Mpc increases to $5 \times 10^{-4}$, corresponding to 0.4 bursts in the reduced 3B sample. We could also argue that the supergalactic concentrations may extend to much larger distances, say 300 Mpc (a redshift of $\sim 0.1$). As pointed out by Shaver & Pierre (18), the distribution of optically selected rich Abell clusters provides evidence for super-galactic structures beyond z = 0.02, and an extension to z = 0.1 appears to be possible (21–23). Galaxy surveys of the nearby ($z \leq 0.2$) universe clearly show large scale power around $100\ h^{-1}$ Mpc 11 and references therein). For the extremely favorable case of such an extended structure combined with a small BATSE volume of $z_{max} = 0.5$ we obtain a fraction of 1.6%, or 14 bursts. It is clear that even an extremely favorable combination of parameters results in a small number of GRB sources that would originate from within the structured volume.

We investigate the flux-ordered sample (3B') of 867 bursts, by selecting the brightest 1/64 (N=14) of these bursts, the top 1/32 (N=27), top 1/16 (N=54), top 1/8 (N=108), top 1/4 (N=217), and top 1/2 (N=433). The total 3B sample only shows deviations of less than $1\sigma$ statistical significance. The reduced sample of 867 bursts with known fluxes shows comparable deviations. Going from the 1/2 sample to the 1/64 sample the quadrupole starts out to be significantly enhanced (2.8 $\sigma$) and then reduces to a value that is consistent with isotropy. The dipole moment for the 1/2 sample is consistent with isotropy and by the time we arrive at the bright samples small anisotropes (1–2 $\sigma$) develop. If these deviations were to reflect the underlying spatial distribution of galaxies located in the superstructure, the quadrupole and dipole moments should be correlated. That this is not the case suggests that the deviations are random and not related to the geometry of the local universe.

**TABLE 1.** Supergalactic statistic of GRBs

| Dataset | N | $F_p^1$ | $D_{SG}$ | $Q_{SG}$ | $\Delta_D$ | $\Delta_Q$ | $B_{SG}$ | $\Delta_B$ |
|---|---|---|---|---|---|---|---|---|
| 3B | 1122 | − | +0.010 | −0.003 | −0.7 | +0.8 | +0.008 | −0.9 |
| 3B' | 867 | 0.05 | −0.003 | +0.006 | −1.3 | +1.6 | +0.006 | −1.0 |
| top 1/2 | 433 | 0.07 | −0.000 | +0.031 | −0.8 | +2.8 | −0.011 | −1.9 |
| top 1/4 | 217 | 1.65 | −0.013 | +0.021 | −0.9 | +1.5 | +0.004 | −0.6 |
| top 1/8 | 108 | 3.45 | −0.046 | +0.050 | −1.2 | +2.1 | −0.017 | −1.1 |
| top 1/16 | 54 | 6.56 | −0.136 | +0.017 | −2.0 | +0.7 | −0.020 | −0.9 |
| top 1/32 | 27 | 12.2 | −0.154 | +0.028 | −1.6 | +0.7 | −0.026 | −0.7 |
| top 1/64 | 14 | 21.0 | −0.228 | +0.000 | −1.6 | +0.1 | +0.048 | +0.4 |

If the sources are not only concentrated near the (0,0) direction, but also in the antipodal direction, the statistic of $\cos\theta_{SG}$ could be diluted because two excesses along opposite directions may partially compensate. The distribution of galaxies within ∼ 100 Mpc resembles this kind of geometry (10). To test the hypothesis that some part of these known structures induces a bipolar anisotropy we use the quadrupole statistic

$$B_{SG} = \langle \cos^2\theta_{SG} \rangle - \frac{1}{3} \qquad (4)$$

The statistical error in this measure is identical to $\sigma_Q$. The bias in B expected from uneven sampling of the sky is +0.016. There is no evidence for a bipolar distribution of GRBs, using $(0,0)_{SG}$ as symmetry axis (Table 1).

The absence of a quadrupole moment in the supergalactic frame can be used to derive a lower limit on the sampling distance (redshift) of BATSE. If we simply model the concentration of sources towards the plane as a Watson distribution with $\kappa = -3$ (which places 1/2 of the bursts within 15.7 degrees of the SG plane; 3), detection of the quadrupole at the 2 $\sigma$ level requires that more than ∼ 10% of all sources are located within $z_0$. For standard candle, non-evolving sources this implies a lower limit on the sampling redshift of $z_{max} \sim 0.065$, i.e., detection to a minimum distances of 200 $h^{-1}$ Mpc.

## CONCLUSIONS

We conclude that there is currently no compelling evidence for a statistically significant concentration of sources in super-galactic coordinates, consistent with expectations for a cosmological burst origin. While BATSE has probably detected giant explosions near the "edge" of the universe, it is not clear how many events may have come from "local" galaxies. The arguably most promising "smoking gun" of cosmological burst models that trace the light is supergalactic anisotropy. So far we have not detected it, but we also did not yet expect it to be detectable (see also 16).

This work was supported in part by NASA under grant NAG 5-1578, and by the Deutsche Forschungsgemeinschaft under grant Ma95/3-1. We dedicate this publication to the memory of Gérard de Vaucouleurs.

## REFERENCES

1. Biermann, P. L., Nucl. Phys, B, **43**, 221 (1995).
2. Briggs, M. S. ApJ **407**, 126 (1993).
3. Briggs, M. S. *et al.*, ApJ, in press (1996).
4. Cohen, E. & Piran, T. ApJ, **444**, L25 (1995).
5. de Vaucouleurs, G. Vistas in Astr., **2**, 1584 (1956).
6. de Vaucouleurs,G., de Vaucoleurs, A., Corwin, H. G., Buta, R. J. Paturel, G. & Fouqué, S. *The Third Reference Catalog of Bright Galaxies*, Springer, (1991).
7. Elbert, J. W. & Sommers, P. ApJ, **441**, 151 (1995).
8. Fishman, G. J. *et al.*, ApJS, **92**, 229 (1994).
9. Fenimore, E. E. *et al.*, Nature, **366**, 40 (1993).
10. Kolatt, T. Dekel, A., & Lahav, O., MNRAS, **275**, 797 (1995).
11. Landy, S. D., *et al.* 1996, ApJ, **456**, L1
12. Meegan, C. A. *et al.*, ApJ, submitted (1995).
13. Milgrom, M. & Usov, V. ApJ, **449**, L37, (1995).
14. Peebles, P. J. E. *Principles of Physical Cosmology* Princeton Univ. Press, (1993).
15. Plunkett, S. P., *et al.*, ApSS, **231**, 271 (1995).
16. Quashnock, J. M., these proceedings (1996).
17. Rutledge, R. E., Hui, L. & Lewin, W. H. G. MNRAS, **276**, 753 (1995).
18. Shaver, P. A. & Pierre, M. A&A, **220**, 35 (1989).
19. Shaver, P. A. Aust. J. Phys., **44**, 759 (1991).
20. Stanev, T. *et al.*, Phys. Rev., **75**, 3056 (1995).
21. Tully, R. B. ApJ, **303**, 25 (1986).
22. Tully, R. B. ApJ, **323**, 1 (1987).
23. Tully, R. B. ApJ, **388**, 9 (1992).
24. Vietri, M. ApJ, **453**, 883 (1995).
25. Waxman, E. Phys. Rev. Lett., **75**, 386 (1995).
26. Waxman, E. ApJ, **452**, L1 (1995).
27. Wickramasinghe, W. *et al.* ApJ, **411**, L55 (1993).